# Cellular-Automata and Innovation within Indonesian Traditional Weaving Crafts

A Discourse of Human-Computer Interaction

Hokky Situngkir
[hs@compsoc.bandungfe.net]
Dept. Computational Sociology
Bandung Fe Institute

**Abstract**

The paper reports the possibility of Indonesian traditional artisans of weaving designs and crafts to explore the cellular automata, a dynamical model in computation that may yield similar patterns. The reviews of the cellular automata due to the perspective of weaving process reveals that the latter would focus on macro-properties, i.e.: the strength of structural construction beside the aesthetic patterns and designs. The meeting of traditional weaving practice and the computational model is delivered and open the door for interesting discourse of computer-aided designs for the traditional artists and designers to come.

**Keywords**: cellular automata, weaving designs, computer-aided designs, Indonesia, crafts.

## 1. Introduction

The vast diversity of Indonesian ethnic groups hold the vast cultural elements. The diversity has emerged in motifs [4, 7], folk songs [5], architectures [6], and more. Each unique cultural elements attach to the respective way of life, everyday routines and customs. One of interesting elements across the traditional cultures is the motifs in weaving crafts and products. Indonesian traditional life use woven objects everyday be it as decorative embellishment or functional domestic devices and tools. The weaving products vary from the form of fabric, dresses, to the kitchen tools, like containers of kitchen objects, mats, and more.

Every places in Indonesia has unique ornaments within their weaving arts, crafts, and products. Nonetheless, the "drawings" ornaments within woven products has great similarities due to the cross-wiring elements of the matters. The drawings are placed upon the lattices and grids thus, relative to the any traditional paintings, like *batik* for instance, the weaving products meet some particular similarities from one another.

On the other hand, there have been many studies of possible patterns yielded computationally from mathematical models [1]. There have been also studies related to incorporating the traditional knowledge of weaving techniques for newly and modern designs of architecture [3]. Among all of the existing models, cellular automaton is an interesting rising computational model that could possibly give new insights for its patterns and emerged visual similarities. Knitting and crochet, embroidery, weaving, and fashion designs are all using 2-dimensional grids as media, and so the emerging patterns generated from the rules of cellular automata.

Introducing cellular automata to the traditional artists and crafters in Indonesia thus become interesting. One of the famous center for weaving traditions and industries in Indonesia is in the Province of Central Kalimantan, Indonesia. A three days' workshop are delivered to introduce the basic concepts of cellular automata while observing how they actually produce the traditional art employing the weaving methods, i.e.: the rattan weaving, traditional fashion *Sasirangan* workers, and traditional knitting and embroidery.

The paper discusses the equivalence between weaving techniques and the cellular automata pattern computational generations and how there seem to be interaction between both. Some interesting aspects of cellular automata is discussed and compared to the one incorporated in the traditional weaving process. As cellular automata is also an important computational model capturing many dynamical systems, the discussion employing the structure and beautiful patterns in traditional weaving methods may reveal some interesting aspects.

## 2. Weaving Automata

Indonesian traditional weaving ornaments are in the 2-dimensional discrete world of grids. Figures and visual aesthetics are there due to placements of states within the lattices and grids. The patterns may come from the macro to micro ornamentation, but it may also come from the micro-rules with the strength yielded by the cross-wiring elements. Resulting structural strength is one thing that is important in any woven patterns. Thus, a woven pattern is not only about information about aesthetics, but also about constructions. This is probably in some places in Indonesia, the explorations of woven motifs are not so vast; even though the collections of all patterns from all over the place within the archipelago are definitely so vast to explore.

The structural patterns of woven products used two cross-wiring elements, the "*lusi*" or "*lungsi*" the elements that is "warped", and the "weft" called "*pakan*". The patterns are emerged from the numbers of warps cross-wired by the wefts. Innovations also may use different colors of weft in order to the accepted aesthetically by the designers and artists. Practically, the elements can be threads, sheets of rattan, or processed leaves of palms or any other thin yet strong enough plants. There are some differences while the weaving is practiced by using the different materials, but the principles are relatively similar.

On the other hand, cellular automata represents a discrete world with parallel dynamics within the microstructures. The two-dimensional patterns are in the form of lattices of crossing vertical and horizontal grids. The emerging macro properties are the patterns yielded by the applied initial conditions and specified rules [12]. Updating rules act as function that evaluates one previous discrete iteration states of lattices surrounding the observed cell. A state of a cell depends upon the states of the neighboring cells. Interacting cells are the emerging the macro-patterns.

While the cellular automata can generally be written as, updating function based on the previous state of neighbors [11],

$$a_i^{t+1} = \psi_k(a_{i-n}^t, \cdots, a_i^t, \cdots, a_{i+n}^t) \tag{1}$$

the practice of weaving generally has more complicated rules of updating. It is not merely the states of previous neighbors that is put into account. It is also possible to evaluate more neighbors over time when designing the weaving motifs. Mathematically, we can write that there is a function,

$$a_i^{t+1} = \psi_{k,w}(a_{i-n}^{t-w}, \cdots, a_i^{t-w}, \cdots, a_{i+n}^{t-w}, \cdots, a_{i-n}^t, \cdots, a_i^t, \cdots, a_{i+n}^t) \tag{2}$$

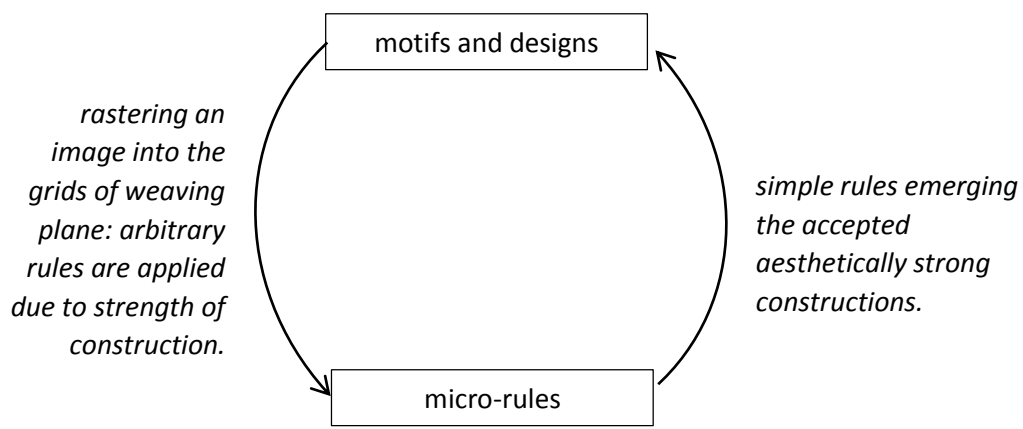

**Figure 1.** The "logic" of weaving patterns

Due to equation (2), when $n = 2$ and $w = 1$, we have exactly the elementary cellular automata. The question now is, is all the 256 rules of the elementary cellular automata can be implemented into weaving techniques? To answer this question, we calculate the entropy of all rules while they are implemented in 50 iterations with random initial conditions.

Entropy, roughly speaking, can be seen as the variables representing the information in the respective evolution of the applied cellular automata rules. We can write the entropy of a set of $Q$ as,

$$H(Q) = -\sum P(q) \log_2 P(q) \qquad (3)$$

summing over the elements of $q \in Q$, where $P(A)$ is the probability of the existence of one state $(A)$ of the cellular automata, and $P(B)$ is in the other opposite states $(B)$. However, due to the construction issue, crossing between state $A$ and $B$ should not be very long otherwise the strength of the weaving would not be able to hold the structure. In other words, the ratio between $P(A)$ and $P(B)$, denoted as $h$, should not be too small or too big,

$$0 \ll h \ll h_{max} \qquad (4)$$

Figure 2 shows the possible relations between the $H(Q)$ and $h$ in the elementary cellular automata. The figure shows that the possible weaving designs should be in the middle of the x-axes, while apparently, the cellular automata rules in the middle is also the ones with highest entropy.

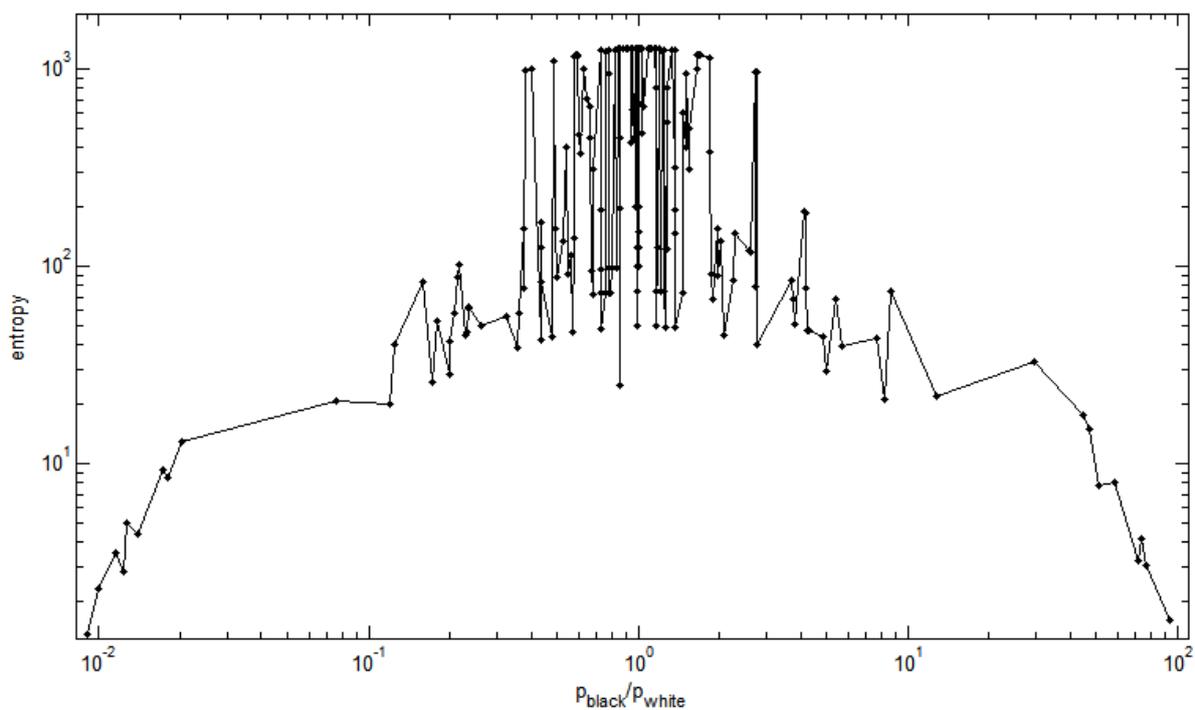

**Figure 2.** The log-log plot of the entropy and ratio between two possible states in elementary cellular automata.

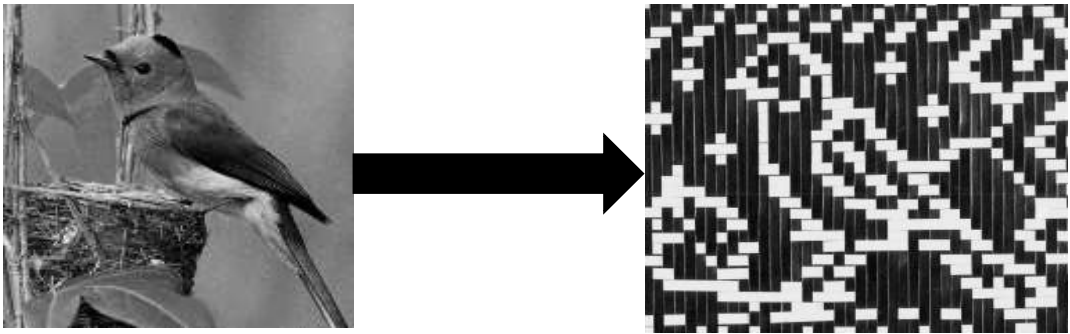

**Figure 3.** The raster process transforming a natural image into the weaving products.

Referring to the discussions brought by Langton [2] on Wolfram's classifications about cellular automata, the most complex structure is always in between the simplest ones, be it very simple or very random. The interesting fact shown in figure 2 depicts the similar aspect; the most complex structure is in between the highest dominance of the *warp* and the *weft*. Balance is important between the two crossing elements of weaving structure, otherwise the structure will not hold as a weaving structure.

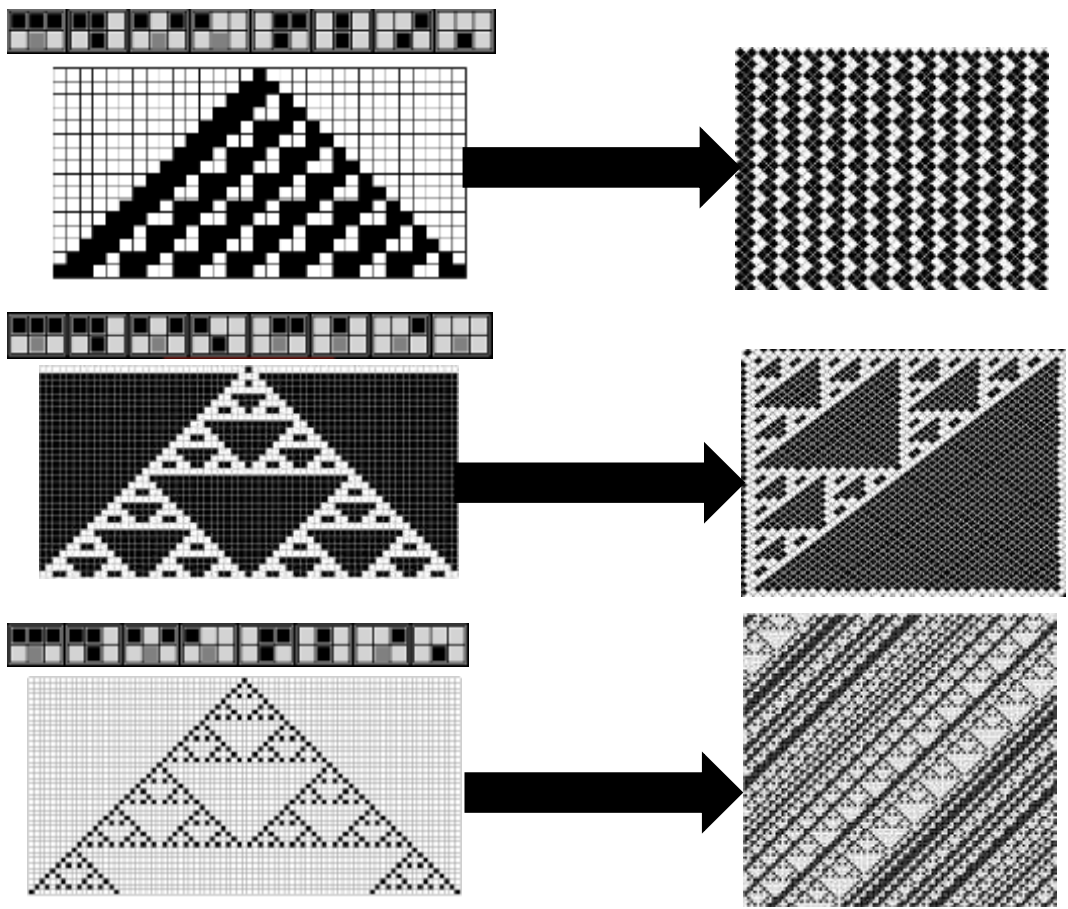

**Figure 4.** Some examples of weaving designs experimented.

## 3. Weaving-experiments with Cellular Automata

A software introducing the adaptation of cellular automata into weaving designs is packed as a computer-aided design tool and tried by traditional artisans in Banjarbaru, South Kalimantan, Indonesia, for us to observe how their background aesthetics meet the exploration of motifs with the cellular automata. The result is interesting for many possible motifs using the simulated automata has never yet merged into their senses, yet they are possible to be implemented.

While the traditional artisans get used to work by using their own standard and "rules" designing the woven craft, the experiments with the cellular automata has made it is possible to explore more patterns that is possible for implementation. Most of the traditional designs were actually in between the "tail" and the "head" of rule-space depicted in figure 2. Explorations with characteristics like those in the "head" of the rule-space, mostly delivered by using the raster processes as shown in figure 3.

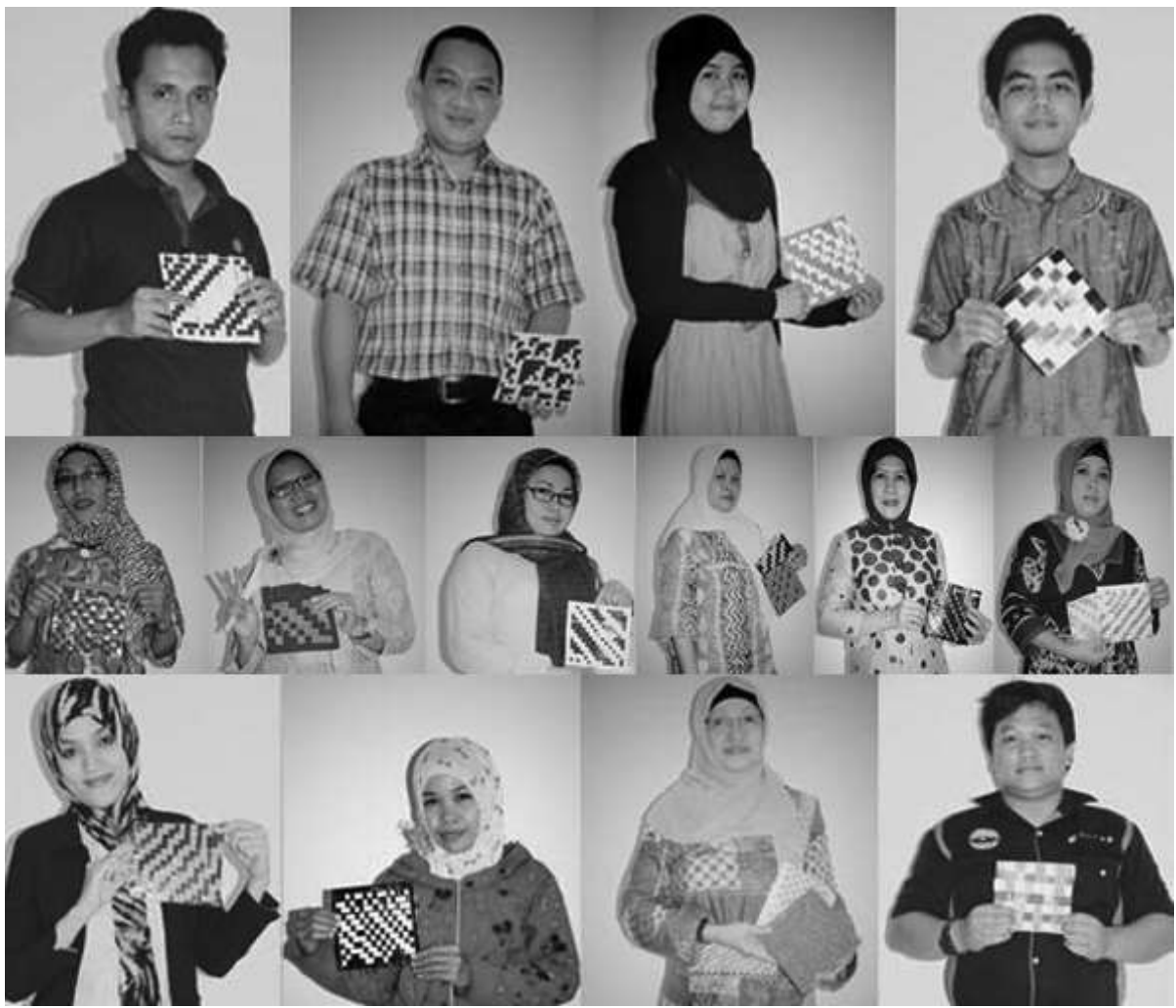

**Figure 5.** Some examples of weaving designs experimented.

Figure 5 shows their working project of weaving by the implementation of the cellular automata with various rules. The implementation to the actual products, e.g.: bags, mats, to dresses is waiting ahead.

## 4. Concluding Remarks

The review of cellular automata in the perspective of Indonesian traditional weaving practice is demonstrated. This is delivered by observing the similar aspects of the simple elementary cellular automata and the hypothetical rule-making of traditional weaving design process. While cellular automata is observed mostly for its possible emerging macro-dynamics, the weaving processes put the structural construction into account while making the pattern designs.

Explorations by using all possible rules with the packed computer-aided designs adapting the cellular automata has made the traditional artisans grasps new possible designs expected to enrich their future patterns designs. Further works will be more about the observations to the implementation of more explorations of various cellular automata in more traditional weave-able materials.

**Acknowledgement**

Author thank Ministry of Research & Technology of the Republic of Indonesia holding the workshop discussed in the paper, the participation of the traditional artists and crafters from Banjarbaru, Central Kalimantan, and the Invasa Co. as the committee of the workshop.